\documentclass[twocolumn,showpacs,preprintnumbers,amsmath,amssymb]{revtex4}
\usepackage{graphicx}% Include figure files
\usepackage{dcolumn}% Align table columns on decimal point
\usepackage{bm}% bold math
\begin{document}
%\preprint{APS/123-QED}
\title{Torsion cosmological dynamics}% Force line breaks with \\
\author{Xin-zhou Li} \email{kychz@shnu.edu.cn}
% \altaffiliation{Shanghai United Center for Astrophysics, Shanghai Normal University 100 Guilin Road, Shanghai 200234, China}%Lines break automatically or can be forced with \\
\author{Chang-bo Sun}%
\author{Ping Xi}
\affiliation{Shanghai United Center for Astrophysics(SUCA),
 Shanghai Normal University, 100 Guilin Road, Shanghai 200234,China}
%\homepage{http://www.Second.institution.edu/~Charlie.Author}
%\affiliation{
%Second institution and/or address\\
%This line break forced% with \\
%}%
\date{\today}% It is always \today, today,
             %  but any date may be explicitly specified
\begin{abstract}
In this paper, the dynamical attractor and heteroclinic orbit have
been employed to make the late-time behaviors of the model
insensitive to the initial condition and thus alleviate the
fine-tuning problem in the torsion cosmology. The late-time de
Sitter attractor indicates that torsion cosmology is an elegant
scheme and the scalar torsion mode is an interesting geometric
quantity for physics. The numerical solutions obtained by Nester et
al. are not periodic solutions, but are quasi-periodic solutions
near the focus for the coupled nonlinear equations.
\end{abstract}
% PACS, the Physics and Astronomy
                             % Classification Scheme.
%\keywords{Suggested keywords}%Use showkeys class option if keyword
                              %display desired
\maketitle
\section{INTRODUCTION}
The current observations, such as SNeIa (Supernovae type Ia), CMB
(Cosmic Microwave Background) and large scale structure, converge on
the fact that a spatially homogeneous and gravitationally repulsive
energy component, referred as dark energy, accounts for about $70$
\% of the energy density of universe. Some heuristic models that
roughly describe the observable consequences of dark energy were
proposed in recent years, a number of them stemming from a certain
physics \cite{Padmanabhan01} and the others being purely
phenomenological \cite{Copeland02}. About thirty years ago, the
bouncing cosmological model with torsion was suggested in
Ref.\cite{Kerlick}, but the torsion was imagined as playing role
only at high densities in the early universe. Goenner et al. made a
general survey of the torsion cosmology \cite{Goenner}, in which the
equations for all the PGT (Poincar{\'e} Gauge Theory of gravity)
cases were discussed although they only solved in detail a few
particular cases. Recently some authors have begun to study torsion
as a possible reason of the accelerating universe \cite{Boeheretal}.
Nester and collaborators \cite{shie03} consider an accounting for
the accelerated universe in term of PGT: dynamic scalar torsion.
With the usual assumptions of homogeneity and isotropy in cosmology,
they find that torsion field could play a role of dark energy. This
elegant model has only a few adjustable parameters, so scalar
torsion may be easily falsified as "dark energy".

The fine-tuning problem should be one of the most important issues
for the cosmological models, and a good model should limit the
fine-tuning as much as possible. The dynamical attractor of the
cosmological system has been employed to make the late-time
behaviors of the model insensitive to the initial condition of the
field and thus alleviates the fine-tuning problem \cite{Hao04}. In
this paper, we study  attractor and heteroclinic orbit in the
torsion cosmology. We show that the late-time de Sitter behaviors
cover a wide range of the parameters. This attractor indicates that
torsion cosmology is an elegant scheme and the scalar torsion mode
is an interesting geometric quantity for physics. Furthermore, there
are only exact periodic solutions for the linearized system, which
just correspond to the critical line (line of centers). The
numerical solutions in Ref.\cite{shie03} are not periodic, but are
quasi-periodic solutions near the focus for the coupled nonlinear
equations.

\section{AUTONOMOUS EQUATIONS}
PGT \cite{Hehl05} based on a Riemann-Cartan geometry, allows for
dynamic torsion in addition to curvature. The affine connection of
the Riemann-Cartan geometry is
\begin{equation}\label{PGT}
  \Gamma_{\mu\nu}{}^\kappa=\overline{\Gamma}_{\mu\nu}{}^\kappa+\frac{1}{2}(T_{\mu\nu}{}^\kappa+T^\kappa{}_{\mu\nu}
  +T^\kappa{}_{\nu\mu})\,,
\end{equation}
where $\bar{\Gamma}_{\mu\nu}{}^{\kappa}$ is the Levi-Civita
connection and $T_{\mu\nu}^{\kappa}$ is the torsion tensor.
Meanwhile, the Ricci curvature and scalar curvature can be written
as
\begin{eqnarray}
   &&R_{\mu\nu} = \overline{R}_{\mu\nu} + \overline{\nabla}_\nu T_\mu +\frac{1}{2}
   (\overline{\nabla}_\kappa - T_\kappa)(T_{\nu\mu}{}^\kappa+T^\kappa{}_{\mu\nu}+T^\kappa{}_{\nu\mu})\nonumber\\
   &&+\frac{1}{4}(T_{\kappa\sigma\mu}T^{\kappa\sigma}{}_\nu+2T_{\nu \kappa \sigma}T^{\sigma \kappa}{}_\mu)\,,\\
   &&R=\overline{R} + 2\overline{\nabla}_\mu T^\mu+\frac{1}{4}(T_{\mu\nu \kappa}T^{\mu\nu \kappa}
    +2T_{\mu\nu \kappa}T^{\kappa\nu\mu}-4T_\mu T^\mu),\nonumber\\
\end{eqnarray}
where $\bar{R}_{\mu\nu}$ and $\bar{R}$ are the Riemannian Ricci
curvature and scalar curvature, respectively, and $\bar{\nabla}$ is
the covariant derivative with the Levi-Civita connection and
$T_{\mu}\equiv T^{\nu}_{\mu\nu}$. According as Ref.\cite{shie03} we
take the restricted form of torsion in this paper
\begin{eqnarray}
T_{\mu\nu\rho}=\frac{2}{3}T_{[\mu}g_{\nu]\rho}
  \label{restrictedT}
\end{eqnarray}
therefore, the gravitational Lagrangian density for the scalar mode
is (For a detailed discussion see Ref.\cite{Nester2})
\begin{eqnarray}
  L_g &=& -\frac{a_0}{2}R +\frac{b}{24}R^2\nonumber\\
        &&+\frac{a_1}{8}(T_{\nu\sigma\mu}T^{\nu\sigma\mu}
         +2T_{\nu\sigma\mu}T^{\mu\sigma\nu}-4T_\mu
         T^\mu)\,,\label{Lg0+mode}
\end{eqnarray}
Since current observations favor a flat universe, we will work in
the spatially flat Robertson-Walker metric. According to the
homogeneity and isotropy, the torsion $T_{\mu}$ should be only time
dependent, so one can let $T_{t}(t)\equiv\Phi(t)$ and the spatial
parts vanish since we have taken the restricted form
(\ref{restrictedT}) of torsion. For the general form, the torsion
tensor have two independent components
\cite{Goenner}-\cite{Boeheretal}. From the field equations one can
finally give the necessary equations for the matter-dominated era to
integrate (For a detailed discussion see Ref.\cite{shie03})
\begin{eqnarray}
      \dot{H}&=&\frac{\mu}{6a_1}R-\frac{\rho}{6a_1}-2H^2\,,\label{dtH}\\
      \dot{\Phi}&=&-\frac{a_0}{2a_1}R-\frac{\rho}{2a_1}-3H\Phi
                   +\frac{1}{3}\Phi^2\,,\label{dtphi}\\
      \dot{R}&=&-\frac{2}{3}\left(R+\frac{6\mu}{b}\right)\Phi\,,\label{dtR}
\end{eqnarray}
where $\mu= a_1-a_0$ and the energy density of matter component
\begin{eqnarray}
  &&\rho=\frac{b}{18}(R+\frac{6\mu}{b})(3H-\Phi)^2-\frac{b}{24}R^2-3a_1H^2
  \,.\label{fieldrho}
 \end{eqnarray}
One can scale the variables and the parameters as
\begin{eqnarray}
&&t\rightarrow l_{p}^{-2}H_{0}^{-1}t,\,\, H\rightarrow
l_{p}^{2}H_{0} H,
\,\, \Phi\rightarrow l_{p}^{2}H_{0}\Phi,\,\, R\rightarrow l_{p}^{4}H_{0}^{2}R,\nonumber\\
&&a_0\rightarrow l_{p}^{2}a_0,\,\, a_1\rightarrow l_{p}^{2}a_1,\,\,
\mu\rightarrow l_{p}^{2}\mu,\,\, b\rightarrow
l_{p}^{-2}H_{0}^{-2}b,\label{scale}
\end{eqnarray}
where $H_0$ is the present value of Hubble parameter and
$l_p\equiv\sqrt{8\pi G}$ is the Planck length. Under the transform
(\ref{scale}), Eqs. (\ref{dtH})-(\ref{dtR}) remain unchanged. After
transform, new variables $t$, $H$, $\Phi$ and $R$, and new
parameters $a_0$, $a_1$, $\mu$ and $b$ are all dimensionless.
Furthermore, the Newtonian limit requires $a_0=-1$. Obviously, Eqs.
(\ref{dtH})-(\ref{dtR}) is an autonomous system, so we can use the
qualitative method of ordinary differential equations. It is worth
noting that in the analysis of critical points, Copeland et al.
\cite{Copeland} introduced the elegant compact variables which are
defined from the Friedmann equation constraint, but in our case, the
Friedmann equation can not be written as the ordinary form, so the
compact variables are not convenient here. Therefore, we will
analyze the system of Eqs.(\ref{dtH})-(\ref{dtR}) using the
variables $H$, $\Phi$ and $R$ under the transform (\ref{scale}).
\section{LATE TIME DE SITTER ATTRACTOR}
In the case of scalar torsion mode, the effective energy-momentum
tensor can be represented as
\begin{eqnarray}
\widetilde{T}_t{}^t&=&\!\!-3\mu H^2\!+\frac{b}{18}(R+\frac{6\mu}{b})
  (3H-\Phi)^2-\frac{b}{24}R^2,\label{torho}\\
   \widetilde{T}_r{}^r&=&\widetilde{T}_\theta{}^\theta
   =\widetilde{T}_\phi{}^\phi
     ={1\over 3}[\mu(R-\overline{R})-\widetilde{T}_t{}^t]\,,\label{torpre}
\end{eqnarray}
and the off-diagonal terms vanish. The effective energy density
\begin{eqnarray}
\rho_{eff}=\rho+\rho_{T}\equiv
\rho+\widetilde{T}_{tt},\label{effrho}
  \end{eqnarray}
which is deduced from general relativity. $p_{eff}=p_T\equiv
\tilde{T}_r^r$ is an effective pressure, and the effective equation
of state is
\begin{eqnarray}
w_{eff}=\frac{\widetilde{T}_{r}^{r}}{\rho+\widetilde{T}_{tt}}\label{effw}
  \end{eqnarray}
which is induced by the dynamic torsion.

According to equations (\ref{dtH})-(\ref{dtR}), we can obtain the
critical points and study the stability of these points. There are
five critical points $(H_c, \Phi_c, R_c)$ of the system as follows
\begin{eqnarray}
&(\text i) & (0,0,0)\nonumber\\
&(\text{ii}) &\scriptstyle \left(
\left(\frac{3(1+a_{1})}{8}-A\right)\sqrt{B+C},\frac{3}{2}\sqrt{B+C},-\frac{6(1+a_{1})}{b}\right)\nonumber\\
&(\text{iii}) &\scriptstyle \left(-
\left(\frac{3(1+a_{1})}{8}-A\right)\sqrt{B+C},-\frac{3}{2}\sqrt{B+C},-\frac{6(1+a_{1})}{b}\right)\nonumber\\
&(\text{iv}) &\scriptstyle \left(
\left(\frac{3(1+a_{1})}{8}+A\right)\sqrt{B-C},\frac{3}{2}\sqrt{B-C},-\frac{6(1+a_{1})}{b}\right)\nonumber\\
&(\text{v}) &\scriptstyle \left(-
\left(\frac{3(1+a_{1})}{8}+A\right)\sqrt{B-C},-\frac{3}{2}\sqrt{B-C},-\frac{6(1+a_{1})}{b}\right)
\end{eqnarray}
where
$A=\scriptstyle\frac{\sqrt{a_1^2(1+a_1)^3(1+9a_1)}}{8a_1(1+a_1)}$,
$B=\scriptstyle-\frac{(1+a_1)(5+9a_1)}{a_1b}$ and
$C=\scriptstyle-\frac{3\sqrt{a_1^2(1+a_1)^3(1+9a_1)}}{a_1^2b}$.
Consider that the parameter $b$ is associated with the quadratic
scalar curvature term $R^2$, so that $b$ should be positive
\cite{shie03}. Evidently, the critical points $(H_c, \Phi_c, R_c)$
are not real values except (0, 0, 0) in the cases of parameters
$b>0$ and $a_1>0$ or $-1\leq a_1<-1/9$.

\begin{figure}[!htbp]
\centering
\includegraphics[height=1.8in,width=1.8in]{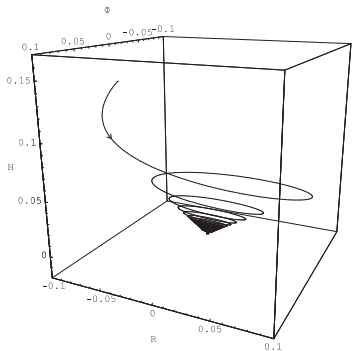}
\caption{The phase diagrams of $(H,\Phi,R)$ for nonlinear system
(\ref{dtH})-(\ref{dtR}) with $a_{1}>0$. We take $a_{1}=0.9,b=8$ and
the initial values $(0.8,0.5,1.1)$. $(0,0,0)$ is an asymptotically
stable focus point.} \label{Focusa1g0}
\end{figure}

 If we consider the linearized
equations, then Eqs. (\ref{dtH})-(\ref{dtR}) are reduced to
\begin{eqnarray}
       \dot{H}=\frac{\mu}{6a_1}R,\qquad  \dot{\Phi}=\frac{1}{2a_1}R,\qquad
      \dot{R}=-\frac{4\mu}{b}\Phi\,.\label{dtHPHIRlinear}
\end{eqnarray}
In the case $a_1>0$, there is only a critical point $(0, 0, 0)$ for
the nonlinear system. Eqs. (\ref{dtHPHIRlinear}) has an exact
periodic solution
\begin{eqnarray}
       &&H=\alpha R_{0}\sin \omega t+\frac{\mu}{3}\Phi_{0}\cos \omega t+H_{0}-\frac{\mu}{3}\Phi_{0},\label{Hps}\\
      &&\Phi=\beta^{-1}R_{0}\sin \omega t +\Phi_{0}\cos \omega t,\label{phips}\\
      &&R=R_{0}\cos\ \omega t-\beta\Phi_{0}\sin \omega t,\label{Rps}
\end{eqnarray}
where $\omega = \sqrt{\frac{2\mu}{a_1b}}$, $\alpha =
\sqrt{\frac{b\mu}{72a_1}}$, $\beta = \sqrt{\frac{8\mu a_1}{b}}$ and
$H_0=H(0)$, $\Phi_0=\Phi(0)$ and $R_0=R(0)$ are initial values.
Obviously, ($H$, $0$, $0$) is a critical line for the linearized
system. However, there is only a critical point $(0, 0, 0)$ for the
nonlinear system which is an asymptotically stable focus. In other
words, there is no periodic solution for the nonlinear system since
the corresponding eigenvalue is ($0$, $-\sqrt{\frac{2\mu}{a_1b}}i$,
$\sqrt{\frac{2\mu}{a_1b}}i$). In Fig.\ref{Focusa1g0}, we plot the
orbits near the point $(0, 0, 0)$ for the nonlinear systems.

Next, the parameters are restricted within $b>0$ and $-1/9\leq
a_1<0$ or $a_1<-1$ in this paper. To study the stability of the
critical points $(H_{c},\Phi_{c},R_{c})$, we write the variables
near the critical points in the form $H=H_{c}+U$, $\Phi=\Phi_{c}+V$,
and $R=R_{c}+X$ with  $U, V, X$ the perturbations of the variables
near the critical points. Substituting the expression into the
system of equations (\ref{dtH})-(\ref{dtR}), one can obtain the
corresponding eigenvalues of critical points (i)-(v)
%\begin{widetext}
\begin{eqnarray}
&&(\text{i})  \scriptstyle(0,-\sqrt{-\frac{2(1+a_{1})}{a_{1}b}},\sqrt{-\frac{2(1+a_{1})}{a_{1}b}})\nonumber\\
&&(\text{ii})  \scriptstyle\left(-\sqrt{B+C},
-3\left(\frac{3(1+a_{1})}{8}-A\right)\sqrt{B+C},
-\left(\frac{1+9a_{1}}{8}-3A\right)\sqrt{B+C}\right)\nonumber\\
&&(\text{iii})  \scriptstyle \left(\sqrt{B+C},
3\left(\frac{3(1+a_{1})}{8}-A\right)\sqrt{B+C},\left(\frac{1+9a_{1}}{8}-3A\right)\sqrt{B+C}\right)\nonumber\\
&&(\text{iv})  \scriptstyle\left(-\sqrt{B-C},
-3\left(\frac{3(1+a_{1})}{8}+A\right)\sqrt{B-C},-\left(\frac{1+9a_{1}}{8}+3A\right)\sqrt{B-C}\right)\nonumber\\
&&(\text{v})  \scriptstyle\left(\sqrt{B-C},
3\left(\frac{3(1+a_{1})}{8}+A\right)\sqrt{B-C},\left(\frac{1+9a_{1}}{8}+3A\right)\sqrt{B-C}\right)
\end{eqnarray}
%\end{widetext}
The properties of the critical points are shown in tables
\ref{cripointsa1l-1} and \ref{cripointsa1l0}. We find that critical
point (ii) is a late time de Sitter attractor in the case of
$-1/9\leq a_1<0$.
\section{NUMERICAL ANALYSIS}
\begin{figure}[!htbp]
\centering
\includegraphics[height=1.4in,width=2.1in]{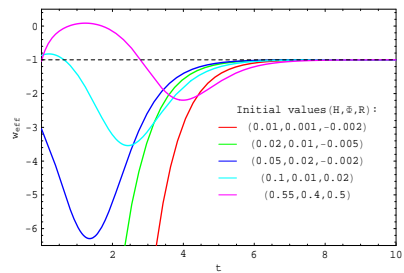}
\caption{The evolution of the equation-of-state parameters $w_{eff}$
for different initial values$(H,\Phi,R)$ with $-1/9 \leq a_{1}<0$.
We take $a_{1}=-1/10,b=8$.}
 \label{Tweffa1}
\end{figure}

In previous sections, we have studied the phase space of a torsion
cosmology. The de Sitter attractor indicates that torsion cosmology
\cite{Kerlick}-\cite{shie03} is an elegant scheme and the scalar
torsion mode is an interesting geometric quantity for physics. In
this section, we study their dynamical evolution numerically.
\begin{table}[thbp]
\caption{The physical properties of critical points for $a_{1}<-1$}
\label{cripointsa1l-1}
\begin{tabular}{c c c c}
\hline
critical points \qquad &property \qquad&$w_{eff}$ \qquad&stability \qquad\\
\hline
(\text{i})& focus & $\pm\infty$ & stable\\
(\text{ii})&saddle&-1&unstable\\
(\text{iii}) &saddle&-1&unstable\\
(\text{iv}) &saddle&-1&unstable\\
(\text{v}) &saddle&-1&unstable\\
\hline
\end{tabular}
\end{table}
The crossing of the $w=-1$ barrier is impossible in the traditional
scalar field models \cite{Caldwell06}. The importance of the torsion
cosmology is further promoted by this impossibility. In Fig.
\ref{Tweffa1}, we plot the dynamical evolution of the equation of
state $w_{eff}$ for different initial values ($H$, $\Phi$, $R$).
Contrary to the quintessence and phantom model \cite{Caldwell06}, the
effective equation of state parameter $w_{eff}$ is dependent on time
that can cross the cosmological constant divide $w_\Lambda=-1$ from
$w_{eff}>-1$ to $w_{eff}<-1$ as the observations mildly indicate.

\begin{figure}[!htbp]
\centering
\includegraphics[height=1.4in,width=2.1in]{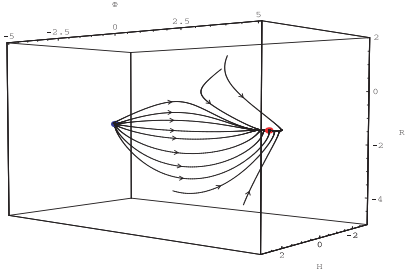}
\caption{The phase diagrams of $(H,\Phi,R)$ with $-1/9 \leq
a_{1}<0$. The heteroclinic orbit connects the critical points case
(iii) to case (ii). We take $a_{1}=-1/10,b=4$.}
\label{heteroclinicorbita1l0}
\end{figure}

Critical points are always exact constant solutions in the context
of autonomous dynamical systems. These points are often the extreme
points of the orbits and therefore describe the asymptotic behavior.
If the solutions interpolate between critical points they can be
divided into a heteroclinic orbit and a homoclinic orbit (a closed
loop). The heteroclinic orbit connects two different critical points
and homoclinic orbit is an orbit connecting a critical point to
itself. In the dynamical analysis of cosmology, the heteroclinic
orbit is more interesting \cite{Li08}. If the numerical calculation
is associated with the critical points, then we will find all kinds
of heteroclinic orbits. Especially, the heteroclinic orbit is shown
in Fig.\ref{heteroclinicorbita1l0}, which connects the positive and
negative attractors.
\begin{table}[thbp]
\caption{The physical properties of critical points for $-1/9\leq
a_{1}<0$} \label{cripointsa1l0}
\begin{tabular}{c c c c}
\hline
critical points \qquad &property \qquad&$w_{eff}$ \qquad&stability \qquad\\
\hline
(\text{i})& saddle & $-\infty$& unstable\\
(\text{ii})&positive attractor&-1&stable\\
(\text{iii}) &negative attractor&-1&unstable\\
(\text{iv}) &saddle&-1&unstable\\
(\text{v}) &saddle&-1&unstable\\
\hline
\end{tabular}
\end{table}
\section{CONCLUSION AND DISCUSSION}
In this paper, we investigate the dynamics of a torsion cosmology,
in which we consider only the "scalar torsion" mode. This mode has
certain distinctive and interesting qualities. We show that the
late-time asymptotic behavior does not always correspond to an
oscillating aspect. In fact, only in the focus case can we declare
that "scalar torsion" mode can contribute a quasinormal oscillating
aspect to the expansion rate of the universe. There are only exact
periodic solutions for the linearized system, which just correspond
to the critical line (line of centers). Via numerical calculation of
the coupled nonlinear equations Nester et al. \cite{shie03} plot
that quasi-periodic solution near the focus.

The late-time de Sitter attractor indicates that torsion cosmology
is an elegant scheme and the scalar torsion mode is an interesting
geometric quantity for physics. We show that the late-time de Sitter
behaviors cover a wide range of the parameters and thus alleviate
the fine-tuning problem. Furthermore, the torsion cosmology has
considered the possibility that the dynamics scalar torsion
(geometric field) could fully account for the accelerated universe,
which is naturally expected from spacetime gauge theory.

\begin{acknowledgments}
This work is supported by National Science Foundation of China grant
No. 10473007 and No. 10671128.
\end{acknowledgments}

\end{document}